\documentclass[journal,draftcls,onecolumn,12pt,twoside]{IEEEtranTCOM}

\usepackage{cite}

\usepackage{graphicx,color,epsfig,rotating,subfigure}
\usepackage{amsfonts,amsmath,amssymb}
\usepackage{algorithm}
\usepackage{subfigure}
\usepackage{algpseudocode}

\long\def\comment#1{}

\newfont{\bbb}{msbm10 scaled 700}

\newfont{\bb}{msbm10 scaled 1100}

\newcommand{\PP}{\mbox{\bb P}}

\newcommand{\bv}{{\bf b}}

\newcommand{\ev}{{\bf e}}

\newcommand{\pv}{{\bf p}}

\newcommand{\uv}{{\bf u}}

\newcommand{\xv}{{\bf x}}
\newcommand{\yv}{{\bf y}}

\newcommand{\zerov}{{\bf 0}}

\newcommand{\Gm}{{\bf G}}

\newcommand{\Ac}{{\cal A}}
\newcommand{\Bc}{{\cal B}}
\newcommand{\Cc}{{\cal C}}
\newcommand{\Dc}{{\cal D}}
\newcommand{\Ec}{{\cal E}}

\newcommand{\Ic}{{\cal I}}

\newcommand{\Lc}{{\cal L}}
\newcommand{\Mc}{{\cal M}}
\newcommand{\Nc}{{\cal N}}
\newcommand{\Oc}{{\cal O}}

\newcommand{\eqdef}{\stackrel{\Delta}{=}}

\newtheorem{definition}{Definition}
\newtheorem{theorem}{Theorem}
\newtheorem{lemma}{Lemma}
\newtheorem{corollary}{Corollary}
\newtheorem{proposition}{Proposition}

\newtheorem{remark}{Remark}

\newtheorem{example}{Example}

\begin{document}

\title{On the Analysis of Puncturing for Finite-Length Polar Codes: \\Boolean Function Approach}

\author{
\IEEEauthorblockN{
              Song-Nam Hong\authorrefmark{1} and Dennis Hui\authorrefmark{2}}\\
\IEEEauthorblockA{\authorrefmark{1}Ajou University, Suwon, Korea,\\
              Email: snhong@ajou.ac.kr}\\
\IEEEauthorblockA{\authorrefmark{2}Ericsson Research, San Jose, CA,\\
              Email: dennis.hui@ericsson.com}
}
\maketitle

%%%%%%%%%%%%%%%%%%%%%
\begin{abstract} 

This paper investigates the impact of puncturing on finite-length polar codes in which a puncturing pattern $\pv^{N}=(p_0,...,p_N)$ is applied to a length-$N$ polar code.. We first introduce two virtual channels to stochastically model the punctured (untransmitted) bits, which are respectively called {\em useless channel model} (UCM) and {\em deterministic channel model} (DCM). Under each model, we derive boolean functions in variables $p_0,...,p_{N-1}$ that can indicate which polarized channels should carry frozen bits. Based on this, we present an efficient method to jointly optimize a puncturing pattern and an information set. Focusing on a fixed information set, we show that there exist the so-called {\em catastrophic} puncturing patterns that will surely lead to a block error and derive their weight distributions recursively. We then propose the two construction methods of a rate-compatible (RC) polar code which ensures that each puncturing pattern in the family is non-catastrophic. Simulation results demonstrate that the proposed RC polar code outperform the RC Turbo code adopted in LTE.

\end{abstract}
%%%%%%%%%%%%%%%%%%


%%%%%%%%%%%%%%%%%%%%%%%%%%%%%%%%%%%%%%%%%%%
\section{Introduction}\label{sec:Intro}

%%%%%%%%%%%%%%%%%%%%%%%%
% 1. Introduction
%%%%%%%%%%%%%%%%%%%%%%%
Polar codes, proposed by Arikan \cite{Arikan2009},  achieve the symmetric capacity of binary-input discrete memoryless channels (BI-DMCs) under a low-complexity successive cancellation (SC) decoder. The finite-length performance of polar codes can be enhanced by using list decoder that enables polar codes to approach the performance of optimal maximum-likelihood (ML) decoder \cite{TalVardy2011}. It was further shown in \cite{TalVardy2011} that a polar code  concatenated with a simple CRC outperforms well-optimized LDPC and Turbo codes even for short lengths. Due to their good performance and low-complexity, polar codes are currently considered for possible deployment in future wireless communication systems (i.e., 5G).

%PUNCTURING

Puncturing is widely used to support various lengths and to construct rate-compatible (RC) codes. In \cite{WangLiu2014, Shin2013, Niu}, a puncturing pattern is produced by a heuristic algorithm and then the information set of the mother polar code is optimized by taking into account the puncturing pattern. Also, an efficient algorithm to jointly optimize a puncturing pattern and the corresponding information set was recently proposed \cite{Miloslavskaya2015}, where an exhaustive search over all possible puncturing patterns is performed, significantly reducing the search-space by using certain symmetry of the polar encoder. It can achieve an optimized frame-error-rate (FER) performance but still require a higher optimization complexity. A low-complexity method for joint optimization was further developed in \cite{Bioglio}. It is remarkable that the above methods cannot be directly used to construct a RC polar code since in this case, all punctured polar codes in the family and the mother polar code should use the identical information set, i.e., the information set cannot be optimized according to puncturing patterns. In \cite{Huawei2014,Saber}, puncturing patterns are optimized for the given information set of the mother polar code. In general, this approach considerably reduces the optimization complexity, at the cost of an increased  FER.

%%%%%%%%%%%%%%%%%%
%  Contributions
%%%%%%%%%%%%%%%%%%

In this paper, we investigate the impact of puncturing on polarized channels of finite-length polar coded and, based on this, we present an efficient method to construct good (rate-compatible) puncturing patterns. First of all, we introduce two {\em virtual} channels that stochastically model punctured bits, which are referred to as useless channel model (UCM) and deterministic channel model (DCM), respectively. On the one hand, in UCM, it is assumed that the output of the virtual channel is independent of its input and hence, the polar decoder is performed by assigning zero log-likelihood ratios (LLRs) for the punctured bits. On the other hand, in DCM, it is assumed that the channel output is equal to its input as a pre-agreed bit (e.g., zero) with probability 1 and hence, the polar decoder is performed by assigning infinite LLRs for the punctured bits. Suppose that a puncturing pattern $\pv^{N}=(p_0,...,p_{N-1})$ is applied to a length-$N$ polar code. 

Our main contributions are summarized as follows.

%%%%%%%%%%%%%%%%%%
\begin{itemize}
\item We derive boolean functions in variables $p_0,...,p_{N-1}$ indicating which polarized channels should be assigned by frozen bits. The index sets of such polarized channels in UCM and DCM are denoted by  $\Dc_{\pv^N}$ and $\Ec_{\pv^N}$, respectively. 
\item We define a {\em reciprocal} puncturing pattern which ensures that $\Bc_{\pv^N}=\Dc_{\pv^N}$ (or $\Bc_{\pv^N}=\Ec_{\pv^N}$), where $\Bc_{\pv^N}$ denote the index set of the locations of punctured bits. Then, we derive a necessary and sufficient condition such that a puncturing pattern is reciprocal. This condition is used to efficiently construct a good puncturing pattern and the corresponding information set.
\item Next, focus on a fixed information set, we show that there exist the so-called {\em catastrophic} puncturing patterns which will surely lead to a block error. Also, we develop an efficient recursive algorithm to characterize those puncturing patterns and their weight distributions. 
\item Based on the above analysis, we present two simple methods to construct a RC polar code which ensures that each puncturing pattern in the family is non-catastrophic. Via simulation results, we demonstrate that the proposed RC polar code can outperform the RC Turbo code adopted in LTE.
\end{itemize}

This paper is organized as follows. In Section~\ref{sec:pre}, we provide some useful notations and definition to be used throughout the paper. 
In Section~\ref{sec:puncturing}, using boolean functions, we analyze the impact of puncturing on the polarized channels of a polar code, and provide an efficient method to jointly optimize a puncturing pattern and an information set. In Section~\ref{sec:CPP}, we define the catastrophic puncturing patterns (which should be avoided) for a given information set and derive their weight distributions. 
In Section~\label{sec:RC}, we propose two simple methods to construct a RC polar code having a family of non-catastrophic puncturing patterns. Simulation results are provided in Section~\ref{sec:NR}.  Section~\ref{sec:con} concludes the paper.

%%%%%%%%%%%%%%%%%%%%%%%%%%%%%%%%%%
%
%   Define a punctured polar code!!!
%
%%%%%%%%%%%%%%%%%%%%%%%%%%%%%%%%%%%%%%%%
\section{Preliminaries}\label{sec:pre}

In this section we provide some useful notations and definitions that will be used in the sequel.

\subsection{Notation}

Let $[a:b]\eqdef\{a,a+1,\ldots,b\}$ for any integers $a$ and $b\geq a$. A polar code of length $N=2^n$ is considered, in which the polarized channels are indexed by $0,1,\ldots,N-1$. We let $\Ac\subseteq [0:N-1]$ denote the information set that contains all the indices of unfrozen-bit channels.  Accordingly, $\Ac^c$ contains all the indices of frozen bit channels. For any $N = 2^n$, let $\Gm_{N} = \Gm_{2}^{\otimes \log(N)}$ be the rate-one generator matrix of all polar codes with blocklength $N$, where $\Gm_{2}$ is the 2-by-2 Arikan Kernel \cite{Arikan2009}. Also, for any $\Mc \subseteq [1:N]$ and $\Nc \subseteq [1:N]$, let $\Gm_{N}(\Mc,\Nc)$ denote the submatrix of $\Gm_{N}$ obtained by selecting rows and columns whose indices belong to $\Mc$ and $\Nc$, respectively.
We define a function $g(\ell): [0:N-1] \rightarrow \{0,1\}^{n}$ which maps $\ell$ onto a binary expansion as
\begin{equation}
g(\ell)=(b_{n},\ldots,b_{1}),
\end{equation} such that $\ell=\sum_{i=1}^n b_i 2^{n-i}$. 
%For a binary sequence $\bv$, $w_{h}(\bv)$ denotes the Hamming weight of $\bv$, i.e., number of ones in $\bv$. We also let $wt(i)=w_{h}(g(i))$. For any two binary sequences $\bv_{1}=(b_{1,n},\ldots,b_{1,1})$ and $\bv_{2}=(b_{2,n},\ldots,b_{2,1})$, 
%%define an operator ``$\prec$" as follows: 
%we say $\bv_{1} \prec \bv_{2}$ if $\{i: b_{1,i}=0\} \subseteq \{i: b_{2,i}=0\}$.
%% $b_{2,i}= 0$ for all $i \in \{j: b_{1,j}=0\}$. 

\subsection{Punctured Polar Codes}\label{subsec:PPC}

In this section, we formally define the polarized channels of a punctured polar code. Let $s$ denote a number of punctured bits. Then, a polar code of blocklength $N$ is punctured by removing a set of $s$ columns from its generator matrix, which has the effect of reducing the codeword length from $N$ to $N_p = N-s$. Formally, a punctured polar code of post-puncturing blocklength $N_p$ is characterized by its ``mother" (unpunctured) polar code of blocklength $N$ and a puncturing pattern $\pv^{N}=(p_0,\ldots,p_{N-1}) \in \{0,1\}^{N}$ with $p_i = 0$ indicating that the $i$-th coded bit is punctured and thus not transmitted. Let $\Bc_{\pv^{N}}=\{i\in [0:N-1] : p_i = 0\}$ be the index set which contains the zero locations in $\pv^N$. Also, we let $\Bc_{\pv^{N}}^c = [0:N-1]\setminus \Bc_{\pv^{N}}$. Then, we have  $N_p = w_h(\pv^{N}) = |\Bc_{\pv^{N}}^c|$.

For a given puncturing pattern $\pv^{N}$, let $\pi_{\pv^{N}}:{\cal Y}^{N} \rightarrow {\cal Y}^{N_p}$ be a projection operator that copies $N_p=w_{h}(\pv^{N})$ coordinates of its input as its output based on the puncturing pattern specified by $\pv^{N}$, where $w_{h}(\pv^{N})$ denotes the number of ones in $\pv^{N}$, i.e. $\yv^{N_{P}} = \pi_{\pv^{N}}(\yv^{N})$ containing the coordinates of $\yv^{N}$ corresponding to the locations of ones in $\pv^{N}$.
The notion of polarized channels in conventional polar codes can be extended to punctured polar codes in a straightforward manner as follows.  For a given (unpunctured) polar code of blocklength $N$ and puncturing pattern $\pv^{N}$, we define the transition probability of the $i$-th polarized channel of the corresponding punctured polar code as
\begin{align}\label{BitChannelPuncturedPolarCode}
& W^{(i)}(\yv^{N_{p}},\uv^{i-1},\pv^{N}|u_i) = \frac{1}{2^{N-1}} \sum_{\uv_{i+1}^{N}} \sum_{\yv^{N} \in \pi^{-1}_{\pv^{N}}\left(\{\yv^{N_{p}}\}\right)} W^{N}(\yv^{N}| \uv^{N}\Gm_{N}),
\end{align}
where $\pi^{-1}_{\pv^{N}}(S) \triangleq \{\yv^{N} \in {\cal Y}^{N}:\pi_{\pv^{N}}(\yv^{N}) \in S\}$ represents the inverse image of $\pi_{\pv^{N}}(\cdot)$ and the channel transition probabilities are
\begin{equation}
W^{N}(\yv^{N}|\xv^{N}) = \prod_{j \in [1:N]} W(y_j|x_j),\label{eq:tran}
\end{equation} where $W(\cdot|\cdot)$ denotes the channel transition probability of the underlying BI-DMC. For the simplicity of notation, we let $W_{\pv^{N}}^{(i)}$ denote the $i$th polarized channel with the transition probability in (\ref{BitChannelPuncturedPolarCode}), and let $I(W_{\pv^{N}}^{(i)})$  denote the corresponding symmetric capacity.

%%%%%%%%%%%%%%%%%%%%%%%%%%%%%%%%%%%%%%
\section{Impact of Puncturing on Polarized Channels}\label{sec:puncturing}

As explained in Section~\ref{subsec:PPC}, punctured coded bits are not transmitted through an underlying channel. Conventionally, a length-$N$ polar decoder is used regardless of the number of punctured bits and thus, the proper LLR values for those bits should be assigned. In this section, we introduce the two {\em virtual} channels that model those punctured bits stochastically, which are respectively called {\em useless} and {\em deterministic} channels. Under the transmitter-receiver agreement, one of the modelings is applied to each punctured bit and according to a chosen model, the proper LLR values are assigned. Then, we analyze their impacts on the polarized channels using boolean functions.

%%%%%%%%%%%%%%%%%%%%%%%%%%%%
\subsection{Stochastic Modelings of Punctured Bits} 

We define the two {\em virtual} channels to stochastically model punctured bits, which are referred to as {\em useless} channel model (UCM) and {\em deterministic} channel model (DCM), respectively. To be specific, they are defined as follows.

\begin{itemize}
\item In UCM, it is assumed that the punctured coded bits are transmitted over a {\em useless} BI-DMC. In this case, a channel output is independent of its input and thus, the corresponding channel can be defined as
\begin{equation}
W(y_j|x_j) = \PP_{Y} (y_j),
\end{equation}for some probability distribution $\PP_{Y}(\cdot)$. Under this model, the LLRs for the punctured coded bits are assigned by zeros since they are assumed to be equally likely of being one or zero.

\item In DCM, it is assumed that the punctured coded bits are transmitted over a {\em deterministic} (or noiseless) BI-DMC. In this case, a channel output is identical to its input with probability 1. In fact, this can be established by assigning the punctured coded bits with some fixed values that have been pre-agreed with the transmitter, so that the encoder restricts the transmitted codewords to be in the {\em offset} subspace specified by the fixed values. In this paper, without loss of generality, the pre-agreed values are assumed to be zeros, i.e., $x_i = 0$ for all $i \in \Bc_{\pv^{N}}$. Accordingly, for an index set $\Ec_{\pv^{N}}$, $\uv_{\Ec_{\pv^{N}}}=(u_i: i \in \Ec_{\pv^{N}})$ should be chosen such that 
\begin{equation}
\uv\Gm_{N}([1:N], \Bc_{\pv^{N}}) = \zerov,
\end{equation}which is equivalent to
\begin{equation}
\uv_{\Ec_{\pv^{N}}} = \uv_{\Ec_{\pv^{N}}^c}\Gm_{N}(\Ec_{\pv^{N}}^c, \Bc_{\pv^{N}})\Gm_{N}(\Ec_{\pv^{N}}, \Bc_{\pv^{N}})^{-1}.\label{eq:cond}
\end{equation} This condition shows that $\uv_{\Ec_{\pv^{N}}}=(u_i: i \in \Ec_{\pv^{N}})$ should not carry information and be used as frozen bits. Also, the condition (\ref{eq:cond}) is satisfied as long as $\Gm_{N}(\Ec_{\pv^{N}}, \Bc_{\pv^{N}})$ is a full-rank. Under this model, 
the LLRs for the punctured coded bits are assigned by $+ \infty$.
\end{itemize}
In the following Sections~\ref{subsec:UCM} and~\ref{subsec:DCM}, we investigate the impact of puncturing on polarized channels under each virtual channel model. In UCM, we show that some polarized channel, determined by a puncturing pattern, should have zero capacity and then identify the index set, denoted by $\Dc_{\pv^{N}}$, of such polarized channels. Namely, we obtain that
\begin{equation}
I(W_{N}^{(i)}) = 0 \mbox{ for } i \in  \Dc_{\pv^{N}}.
\end{equation} Next, in DCM, we identify the index set $\Ec_{\pv^{N}}$ defined above. Definitely, those index sets have an impact on the design of an information set $\Ac$ as in Remark~\ref{remark:1}

\begin{remark}\label{remark:1} The reliabilities of polarized channels of a polar code can be computed by several techniques such as
density evolution under a Gaussian approximation (DE/GA), tracking the mean value, Batthacharyya parameter or mutual information of the Gaussian L-densities \cite{Sarkis}. Let $\Ic=\{i_1,i_2,...,i_N\}$ denote the {\em orderd} index set such that the reliability of polarized channel $i_j$ is not lower than that of polarized channel $i_k$ for $i\geq k$. From this, we can construct the information set 
$\Ac$ of the size $K$, by taking the first $K$ indices from the $\Ic$. Whereas, for a punctured polar code, we need to more carefully construct an information set as follows:

\begin{itemize}
\item In UCM, the information set $\Ac$ is constructed by taking the first $K$ indices from the ordered set  $\Ic \setminus \Dc_{\pv^N}$ where $|\Ic \setminus \Dc_{\pv^N}|=N_p$.
\item Likewise, in DCM, the information set $\Ac$ is constructed by by taking the first $K$ indices from the ordered set $\Ic \setminus \Ec_{\pv^N}$ where $|\Ic \setminus \Ec_{\pv^N}|=N_p$.
\end{itemize}
\end{remark}
\vspace{0.1cm}

\begin{figure}
\centerline{\includegraphics[width=14cm]{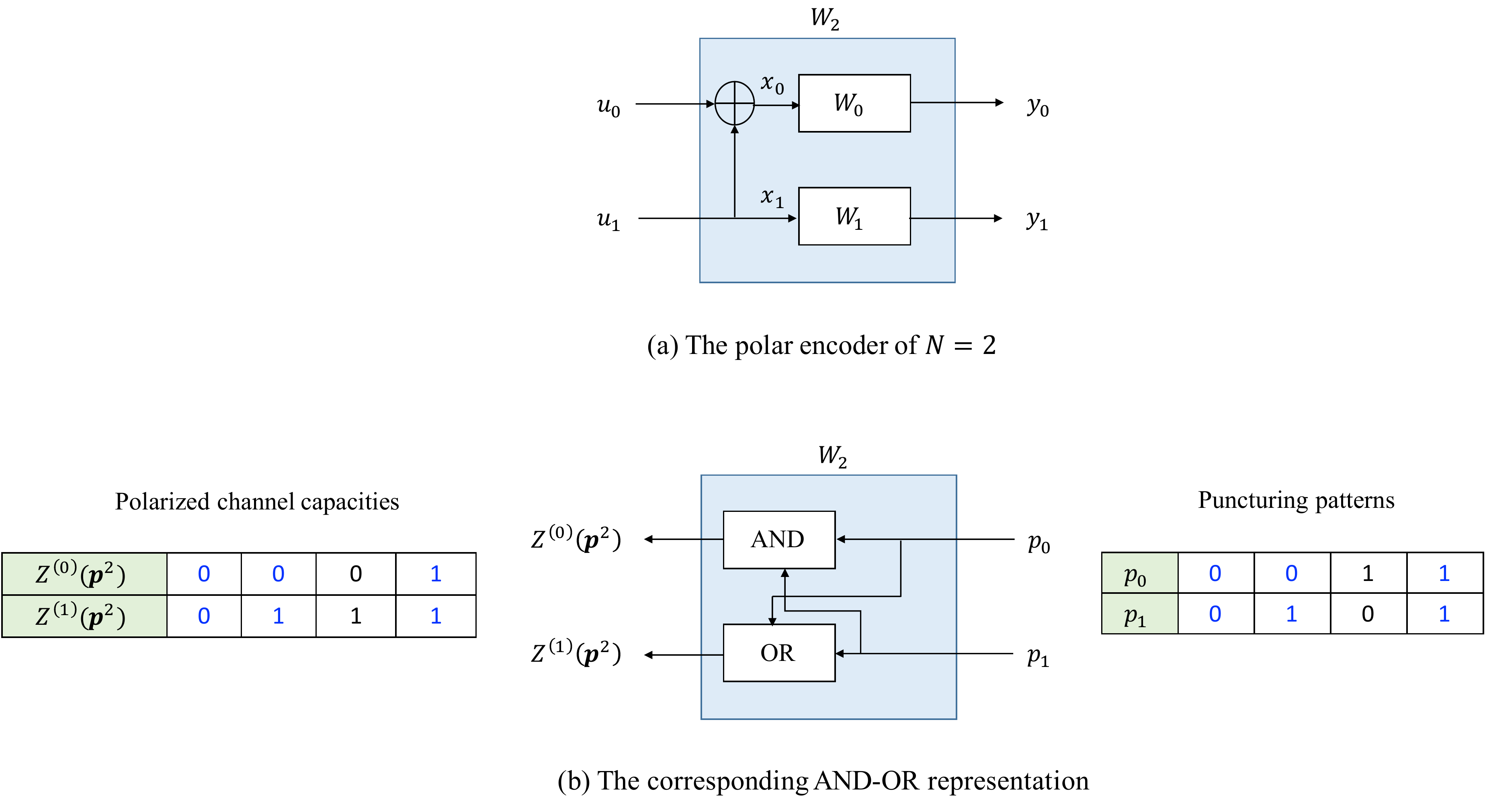}}
\caption{The capacity of polarized channel with a puncturing pattern $\pv^2=(p_{0},p_{1})$ for $N=2$.}
\label{fig:Z}
\end{figure}

\begin{figure}
\centerline{\includegraphics[width=14cm]{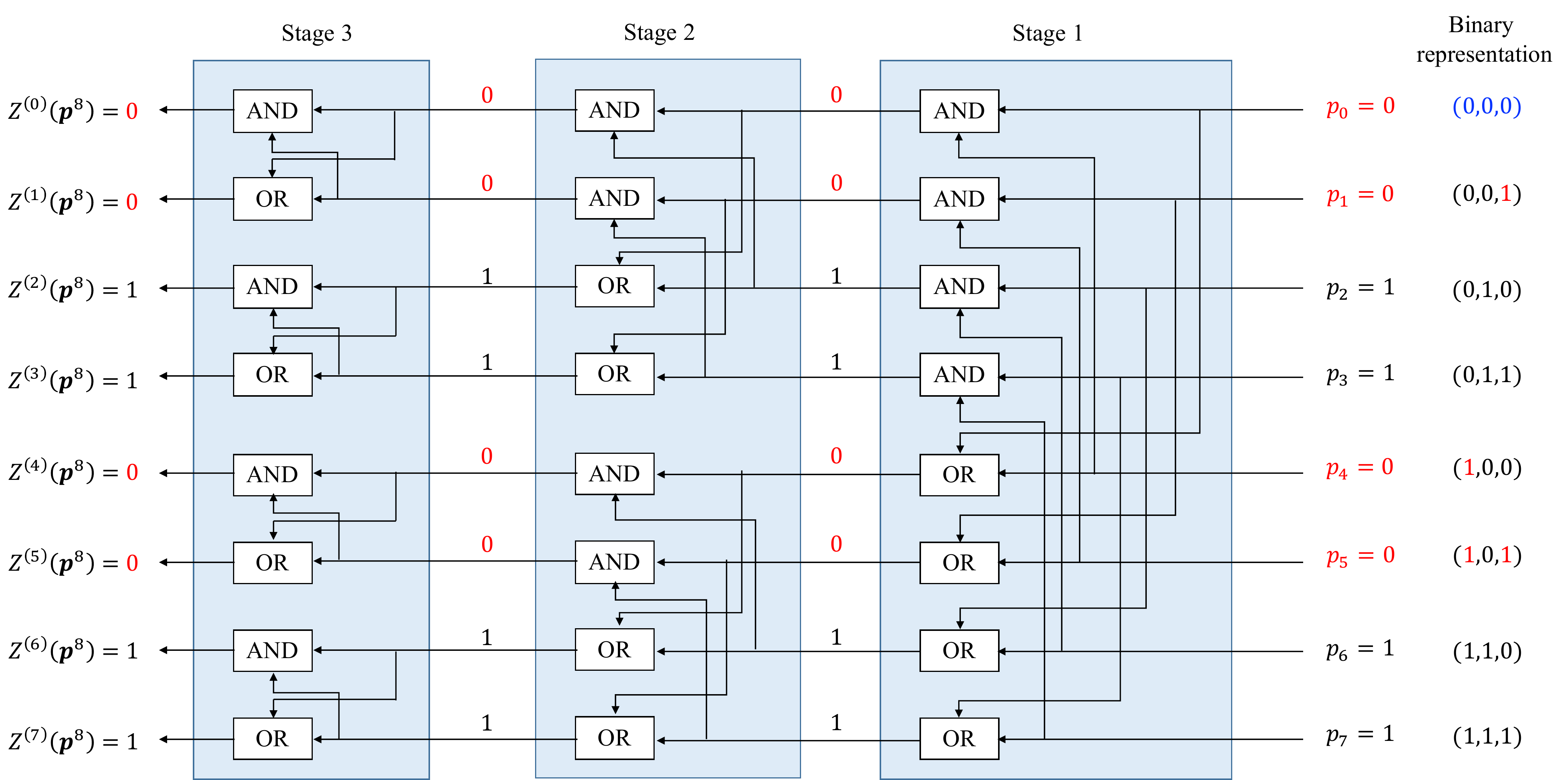}}
\caption{The capacity of polarized channel with a puncturing pattern $\pv^8=(p_{0},p_{1},...,p_{7})$ for $N=8$.}
\label{fig:Z4}
\end{figure}

%{\BLUE
%\begin{definition}{({\em Input parings})} Let $g(i)=(b_1^i,b_2^i,...,b_n^i)$ denote the binary representation of the $i$-th polarized channel. for $i=0,...,N-1$. For each stage $j$, there are $2^{N-1}$ input parings to the corresponding basic models as
%\begin{equation}
%(g(i_1), g(i_2)) \mbox{ with } b_j^{i_1} = 0, b_j^{i_2}=1, \mbox{ and } b_k^{i_1}=b_k^{i_2} \mbox{ for } k\in\{0,..,N-1\}\setminus\{j\},
%\end{equation} for $j\in\{1,...,n\}$.
%\end{definition}
%}

%%%%%%%%%%%%%%%%%%%%%%%%
\subsection{Useless Channel Model (UCM)}\label{subsec:UCM}

Focusing on the perfect underlying channel (i.e., $I(W) = 1$),  we first identify the $\Dc_{\pv^{N}}$ and then show its {\em university} with respect to the underlying B-DMC channel. By incorporating puncturing into the channel, the resulting channel in this case can be either a perfect channel or a completely noisy (i.e. zero-capacity) channel  because a punctured bit is not transmitted. Without loss of generality, it is assumed that $N$ copies of the channel, denoted by $W_i$, $i\in[1:N]$,  are used in which $I(W_i) = 0$ for $i \in \Bc_{\pv^{N}}$ and $I(W_i)=1$ for $i \notin \Bc_{\pv^{N}}$.  We let $\bar{W}_{\pv^N}^{(i)}$ denote the $i$-th polarized channel particularly when the underlying B-DMC is a perfect channel.

We start with the simplest case of $N=2$.  From \cite[Proposition 4]{Arikan2009}, it is easily verified that
\begin{align}
I(\bar{W}_{\pv^2}^{(0)}) &\leq \min\{I(W_{0}),I(W_{1})\}\label{eq:min}\\
I(\bar{W}_{\pv^2}^{(1)}) &\geq \max\{I(W_{0}),I(W_{1})\}.\label{eq:max}
\end{align} When $\pv^2=(1,0)$ is used, the capacities of resulting polarized channels are given by
\begin{align}
I(\bar{W}_{\pv^2}^{(0)}) =0 \;\;\mbox{  and  }\;\; I(\bar{W}_{\pv^2}^{(1)}) =1,
\end{align} because of $I(W_{0})=1$ and $I(W_{1})=0$. From (\ref{eq:min}) and (\ref{eq:max}), we can easily obtain that
\begin{align}
 I(\bar{W}_{\pv^2}^{(0)}) &= \left\{
             \begin{array}{ll}
               0, & \pv^2 \in \{(0,0),(0,1),(1,0)\} \\
               1, & \pv^2 = (1,1)
             \end{array}
           \right.\\
I(\bar{W}_{\pv^2}^{(1)}) &=\left\{
             \begin{array}{ll}
               0, & \pv^2 = (0,0) \\
               1, & \pv^2 \in \{(0,1),(1,0),(1,1)\}.
             \end{array}
           \right.
\end{align} Also, we can simply express $I(\bar{W}_{\pv^2}^{(i)}) $ as a {\em boolean} function in variables $p_{0}$ and $p_{1}$ such as
\begin{align}
I(\bar{W}_{\pv^2}^{(0)}) = p_{0} \wedge p_{1}  \;\;\;\mbox{ and }\;\;\; I(\bar{W}_{\pv^2}^{(1)}) = p_{0} \vee  p_{1}, \label{eq:b1}
\end{align}
where ``$\wedge$" and ``$\vee$" denote the AND and OR operations, respectively, and to simplify notation, we will omit the ``$\vee$" in the following. To simplify the expression and emphasize the boolean function, we let $Z^{(i)}(\pv^N)$ denote the capacity of the $i$th polarized channel, i.e., $Z^{(i)}(\pv^N)=I(\bar{W}_{\pv^N}^{(i)})$.

Note that the boolean functions in (\ref{eq:b1}) can be immediately obtained from Fig.~\ref{fig:Z} (b). The next level of recursion for $N=4$ is shown in Fig.~\ref{fig:Z4}. We first define a binary-output function $h: [0,1]\times [0,1] \rightarrow [0,1]$ as
\begin{equation}
h_{b}(p_0,p_1) = \left\{
             \begin{array}{ll}
               p_0\wedge p_1, & \mbox{ if } b=0 \\
               p_0 p_1, & \mbox{ if } b=1.
           \end{array}
           \right.
\end{equation}  Letting $g(i)=(b_0,b_1)$, we can obtain
\begin{equation}
Z^{(i)}(\pv^4) = h_{b_1}(h_{b_0}(p_0,p_2), h_{b_0}(p_1,p_3)). \label{eq:recursion}
\end{equation} From this, we can obtain the boolean functions for $N=4$ as follows:
\begin{align}
Z^{(0)}(\pv^4) &=p_{0} \wedge p_{1} \wedge p_{2} \wedge p_{3}\label{eq:minimal1}\\
Z^{(1)}(\pv^4) &= p_{0} p_{1} \wedge p_{0}p_{3} \wedge p_1p_2 \wedge p_2p_3 \label{eq:minimal2}\\
Z^{(2)}(\pv^4) &=p_{0} p_{2} \wedge p_{1} p_{3} \label{eq:minimal3} \\
Z^{(3)}(\pv^4) &=p_{0} p_{1} p_{2} p_{3}.\label{eq:minimal4}
\end{align} Repeatedly applying the above recursions as in (\ref{eq:recursion}), we can derive the $Z^{(i)}(\pv^{N}), i=0,\ldots,N-1$ as a function of a puncturing pattern $\pv^{N}$. 
When the perfect underlying channel (i.e., $I(W)=1$) is assumed, we obtain the index set of zero-capacity polarized channels as
\begin{equation}
\Dc_{\pv^{N}}=\{i \in [0:N-1] : Z^{(i)}(\pv^{N}) = 0\}.
\end{equation} Also, the above set is {\em universal} with respect to the underlying B-DMC channel $W$ because  from~\cite[Lemma 4.7]{Korada}, $I(W_{\pv^{N}}^{(i)}) \leq I(\bar{W}_{\pv^{N}}^{(i)})$ for any $W$, and thus $I(W_{\pv^{N}}^{(i)}) = 0$  for any $W$ if $Z^{(i)}(\pv^N) = 0$.

%%%%%%%%%%%%%%%%%%%%%%
\subsection{Deterministic Channel Model (DCM)}\label{subsec:DCM}

Focusing on the simplest case of $N=2$, we can see that the condition (\ref{eq:cond}) is satisfied by choosing 
 $u_0$ and $u_1$ according to a puncturing pattern as
 \begin{align}\label{eq:cases}
 &\left\{
             \begin{array}{ll}
               u_0=0, u_1=0  & \pv^2=(0,0) \\
               u_0=u_1, & \pv^2 = (0,1)\\
               u_1=0 & \pv^2=(1,0).
             \end{array}
           \right.
\end{align} From the above, we can see that in the third case, $u_1$ should be frozen bits, and in the second case, either $u_0$ or $u_1$ should be frozen bits. Since the second polarized channel has a higher capacity than the first polarized channel, we choose $u_0$ as frozen bit in the second case. Let $\bar{\pv}^N$ denote the 1's complement of $\pv^N$. 
Likewise the case of UCM, we can express the (\ref{eq:cases}) using boolean function as
\begin{equation}
Z^{(0)}(\bar{\pv}^2) =\bar{p}_0 \wedge \bar{p}_1 \text{ and } Z^{(1)}(\bar{\pv}^2) = \bar{p}_0 \bar{p}_1,
\end{equation} and the condition (\ref{eq:cond}) is satisfied if $u_i$ is chosen as frozen bit when  $Z^{(i)}(\bar{\pv}^2) = 1$. From this, we obtain the equivalent result with the (\ref{eq:cases}) as follows:
\begin{align}
 Z^{(0)}(\pv^2) &= 1 \mbox{ if } \pv^2 \in \{(0,0)\}\\
 Z^{(1)}(\pv^2) &= 1 \mbox{ if } \pv^2 \in \{(0,1),(1,0),(1,1)\}.
\end{align} In this simplest case, thus, we can identify the 
\begin{equation}
\Ec_{\pv^2} = \{ i \in [0:1]: Z^{(i)}(\bar{\pv}^2) = 1\}.
\end{equation} Following the same recursion procedures in UCM, we can obtain the
\begin{equation}
\Ec_{\pv^{N}}=\{i \in [0:N-1] : Z^{(i)}(\bar{\pv}^{N}) = 1\}.\label{eq:E}
\end{equation} 

\begin{remark} In UCM, the frozen bits $u_{i}$ for $i \in \Dc_{\pv^N}$ can be set by zero, without loss of performance. In DCM, however, the frozen bits $u_{i}$ for $i \in \Dc_{\pv^N}$ should be determined as a binary addition of other bits, and thus, we require another computations (i.e., encoding) to determine the frozen bits. In the next section, it will be shown that for some class of puncturing patterns, those frozen bits can be simply chosen as zeros.
\end{remark}

\begin{example} Consider the case of $N=4$. From (\ref{eq:minimal1})-(\ref{eq:minimal4}), we obtain the corresponding boolean functions as
\begin{align}
Z^{(0)}(\bar{\pv}^4) &=\bar{p}_{0} \wedge \bar{p}_{1} \wedge \bar{p}_{2} \wedge \bar{p}_{3}\\
Z^{(1)}(\bar{\pv}^4) &=\bar{p}_{0} \bar{p}_{1} \wedge \bar{p}_{0} \bar{p}_{3} \wedge \bar{p}_{1}\bar{p}_{2} \wedge \bar{p}_{2} \bar{p}_{3}\\
Z^{(2)}(\bar{\pv}^4) &= \bar{p}_{0} \bar{p}_{2} \wedge \bar{p}_{1}\bar{p}_{3}\\
Z^{(3)}(\bar{\pv}^4) &=\bar{p}_{0} \bar{p}_{1} \bar{p}_{2} \bar{p}_{3}.
\end{align} For $\pv^{4} = (1,0,1,0)$, we have $\Ec_{\pv^{4}=(1,0,1,0)} =\{1,3\}$ from (\ref{eq:E}), which implies that 
the condition (\ref{eq:cond}) is satisfied by assigning $u_2$ and $u_3$ as frozen bits. This is verified as
\begin{equation}
\Gm_4(\{1,3\},\{1,3\})  = \left[ {\begin{array}{cc}
     1 & 0 \\
   1 & 1 \\
  \end{array} } \right]
\end{equation}   is a full-rank.
\end{example}

%%%%%%%%%%%%%%%%%%%%%%%%%%
\subsection{Reciprocal Puncturing Patterns}\label{subsec:RPP}

In Sections~\ref{subsec:UCM} and~\ref{subsec:DCM} above, we showed that for each $\pv^N$, there exists the corresponding index sets  $\Dc_{\pv^{N}} \subseteq \Ac^c$ and $\Ec_{\pv^{N}} \subseteq \Ac^c$ for UCM and DCM, respectively. In order to construct a good information set, they should be identified. In this section, we show that for some puncturing patterns, we can ensure that
 $\Bc_{\pv^N} = \Dc_{\pv^{N}}$ (or $\Bc_{\pv^N} = \Ec_{\pv^{N}}$). Formally, we define:
%%%%%%%%
\begin{definition} A puncturing pattern $\pv^N$ is referred to as {\em reciprocal} if $\Bc_{\pv^N} = \Dc_{\pv^{N}}$ under UCM (or $\Bc_{\pv^N} = \Ec_{\pv^{N}}$ under DCM). 
\end{definition}
\vspace{0.1cm}

Using this, we provide the main theorems of this section below:

%%%%%%%%%%%%%%%%%%%%%%%%%%%%%%%%%%%%%%%%%%%%%%%%%%%%%%%%%%%%%
\begin{theorem}\label{thm1} In UCM, a puncturing pattern $\pv^N$ is reciprocal if and only if  the following properties are satisfied:
\begin{align}
&\mbox{{\bf zero-inclusion property: }} 0 \in \Bc_{\pv^N},\\
&\mbox{{\bf one-covering property: } } \mbox{ if } i \in \Bc_{\pv^N} \mbox{ and } i\succeq_{1} j, \mbox{ then } j \in \Bc_{\pv^N},
\end{align} where $ i\succeq_{1} j$ means that for every digit of '1' in the binary representation of index $j$, the corresponding digit in the index $i$ must also be '1' and $i \succeq_{1} 0$ for every  $i > 0$.
\end{theorem}
\begin{IEEEproof} We first prove sufficiency and then prove necessity.

{\em a) Proof of sufficiency:} It suffices to show that the input and output sequences remain the same after each polarization stage in the  modified encoder with AND-OR operators in Fig.~\ref{fig:Z} (b). Let the binary representation of an index $i$ be denoted by $g(i)=(b_1^i,b_2^i,...,b_n^i)$ where $n = \log_{2}N$. At each polarization stage $k$ for $k\in\{1,...,n\}$, there are $2^{n-1}$ pairs of basic AND-OR operators in the form of that in Fig.~\ref{fig:Z} (b), each of which has the input pairs whose indices correspond to $\ell_0=(b_1^i,...,b_{k-1}^i,0,b_{k+1}^i,..,b_{n}^i)$ and $\ell_1=(b_1^i,...,b_{k-1}^i,1,b_{k+1}^i,..,b_{n}^i)$. The input and the output of each pair are identical if $(p_{\ell_0},p_{\ell_1}) \in \{(0,0),(0,1),(1,1)\}$ as in Fig.~\ref{fig:Z} (b). If the zero-inclusion and one-covering properties are satisfied, then the input and output of each pair of these AND-OR operators are identical. Since this argument holds for every polarization stage, it follows that those properties imply reciprocity.

{\em b) Proof of necessity:} Note that if the input and the output of the modified polar encoder are the same (i.e., $\Bc_{\pv^N} = \Dc_{\pv^N}$), then the input and output after each polarization stage must be also the same because the largest bit index with value '0' is monotonically non-increasing after each polarization stage as the output of the AND operator should be smaller than that of the OR operator for the same input. Now suppose there exist $i$ and $j$ such that $i \in \Bc_{\pv^N}$ and $i \succeq_{1} j$ but $j \notin \Bc_{\pv^N}$. There must exist $\ell_1,\ell_2 \in [1:N]$ such that $i \succeq_{1} \ell_1 \succeq_{1} \ell_2 \succeq_{1} j$, $\ell_1 \in \Bc_{\pv^N}$, $\ell_2 \notin \Bc_{\pv}^N$, and that $\ell_1$ and $\ell_2$ differ only in one bit (say, $k$-th bit), where $k \in \{1,2,...,n\}$. It follows that $b_k^{\ell_1}$, $b_{k}^{\ell_2}=0$, and $b_{m}^{\ell_1}=b_{m}^{\ell_2}$ for all $m\neq k$. Since $\ell_1 \in \Bc_{\pv^N}$ and $\ell_2 \notin \Bc_{\pv^N}$, the input to the pair of AND-OR operators at stage $k$ corresponding to $\ell_1$ and $\ell_2$ should be different from its output as in the 3rd column of the table in Fig.~\ref{fig:Z} (b). It follows that the input and output after (at least) the $k$-th polarization stage must be different and thus the input and the output of the whole modified polar encoder must be different. Thus, $\Bc_{\pv^N} \neq \Dc_{\pv^N}$.
\end{IEEEproof}

%%%%%%%%
\begin{theorem}\label{thm2} In DCM, a puncturing pattern $\pv^N$ is reciprocal if and only if  the following properties are satisfied:
\begin{align}
&\mbox{{\bf $N$-inclusion property:} } N\in \Bc_{\pv^N},\\
&\mbox{{\bf zero-covering property:} } \mbox{ if } i \in \Bc_{\pv^N} \mbox{ and } i\succeq_{0} j, \mbox{ then } j \in \Bc_{\pv^N},
\end{align} where $ i\succeq_{0} j$ means that for every digit of '0' in the binary representation of index $j$, the corresponding digit in the index $i$ must also be '0'.
\end{theorem}
\begin{IEEEproof} The proof follows the same procedures in the proof of Theorem 1.
\end{IEEEproof}

%%%%%%%%
\begin{corollary}\label{cor} Suppose DCM is assumed. If a puncturing pattern $\pv^N$ is reciprocal, then we have
\begin{equation}
u_{i} = 0 \mbox{ for } i \in \Ec_{\pv^N}.
\end{equation}
\end{corollary}
\begin{IEEEproof} Using zero-covering property in Theorem~\ref{thm2}, we define
\begin{equation}
\Phi_N(i)=\{j\in [0:N-1]: i  \succeq_{0} j\} \cup \{N-1\},
\end{equation} for any $i \in [0:N-1]$. Due to the particular construction of $\Gm_{N} = \Gm_{2}^{\otimes \log(N)}$, we can see that $x_i$, $i$-th coded bit, is determined as an addition of $u_i$'s with $i \in \Phi_N(i)$. Since the $\pv^N$ is reciprocal by Hypothesis assumption, we know that $\Phi_N(i) \subseteq \Ec_{\pv^N}$ for  any $i \in \Ec_{\pv^{N}}$. Thus, if we choose $u_i = 0$ for all $i \in \Ec_{\pv^{N}}$, $x_{i} = 0$ for all $i \in \Bc_{\pv^N}$. This completes the proof.
\end{IEEEproof}
\vspace{0.2cm}

From now on, we will prove a class of reciprocal puncturing patterns. 

\begin{definition} Let $\Pi_n$ denote the set of all possible permutation of $(1,2,...,n)$. For any $\pi \in \Pi$, $\pi(i)$ denote the $i$-th element of $\pi$. For any length-$n$ binary vector $\bv=(b_1,...,b_n)$, we define a bit-permutation $\Psi_{\pi}$ with seed vector $\pi \in \Pi_n$ as
\begin{equation}
\Psi_{\pi}(\bv) = (b_{\pi(1)}, b_{\pi(2)},...,b_{\pi(n)}).
\end{equation}
\end{definition}

\begin{proposition}\label{prop} Under UCM, suppose that $\pv^N$ is reciprocal. Then, for any $\pi \in \Pi_n$, the puncturing pattern $\pv_{\pi}^{N}$ with zero-location set $\Bc_{\pv_{\pi}^N}$ is also reciprocal, where
\begin{equation}
\Bc_{\pv_{\pi}^N} =\left\{g^{-1}\left(\Psi_{\pi}(g(i))\right): i \in \Bc_{\pv^N}\right\}.
\end{equation}  The above statement is also hold under DCM.
\end{proposition}
\begin{IEEEproof} The proof follows the fact that one-covering property (or zero-covering property) definitely holds for any bit-permutation $\pi$.
\end{IEEEproof}

%\begin{definition} Let $\Rc_N=\{i_1,i_2,...,i_N\}$ denote the {\em ordered} index set.  Also, let $\Rc_{N}^{(m)}$ denote the subset of $\Rc_{N}$ by taking the first $m$ elements of $\Rc_{N}$. Then, we call $\Rc_{N}$ {\em reciprocal set} if, under UCM,  the corresponding puncturing pattern  $\pv^{N}$ with $\Bc_{\pv^N}=\Rc_{N}^{(m)}$ is reciprocal for any $m\leq N$.
%\end{definition}

%%%%%%%%%%%%%%%%%%%%%%%%%%%%%%%%%
\begin{example} In this example, we provide some reciprocal puncturing patterns. First of all, we have:
\begin{itemize}
\item In UCM, the puncturing pattern $\pv^N$ with $\Bc_{\pv^N}=\{0,1,...,s-1\}$ is obviously reciprocal.
\item In DCM, the puncturing pattern $\pv^N$ with $\Bc_{\pv^N}=\{N-s+1,N-s+2,...,N\}$ is obviously reciprocal.
\end{itemize} Using the above puncturing patterns and from Proposition~\ref{prop}, we are able to generate several reciprocal puncturing patterns. In particular, when $\pi=(n,n-1,...,1)$ (called bit-reverse permutation), the corresponding puncturing patterns in UCM and DCM are called quasi-uniform puncturing (QUP) and reverse QUP (RQUP), respectively. Fig.~\ref{fig2} shows the frame-error-rate (FER) performances of punctured polar codes where QUP and RQUP are used for UCM and DCM, respectively. In this example, we observe that UCM approach is slightly better than DCM approach at lower rate and vice versa at higher rate.
\end{example}
\vspace{0.2cm}

%%%%%%%%
%\begin{example}\label{ex:QUP} Let $\psi_N$ be the bit-reverse permutation of $[0:N-1]$. For example, $\psi_8(3)=6$ as $g(3) = (0,1,1)$ and $g^{-1}((1,1,0)) = 6$. With this definition, we have:
%\begin{itemize}
%\item In UCM, the puncturing pattern with $\Bc_{\pv^N} = \{\psi_N([0:s-1])\}$, called a quasi-uniform puncturing (QUP), is reciprocal.
%\item In DCM, the puncturing pattern with $\Bc_{\pv^N} = \{\psi_N([N-s+1:N])\}$, called a reverse QUP (RQUP), is reciprocal. Also, from Corollary~\ref{cor}, all-zero frozen bits can be used.
%\end{itemize}  Fig.~\ref{fig2} shows the frame-error-rate (FER) performances of punctured polar codes where QUP and RQUP are used for UCM and DCM, respectively. In this example, we observe that UCM approach is slightly better than DCM approach at lower rate and vice versa at higher rate.
%\end{example}
%\vspace{0.1cm}

%%%%%%%%%%%%%%%%%%%%%%%%%%%%%%%%%%%%%%%%%%%%%%%%%%%%%%%%%%%%%%%%%%%%%%%%%%%%%%%
%%%%%%%%%%%%%%%%%%%%%
\begin{figure}
\centerline{\includegraphics[width=14cm]{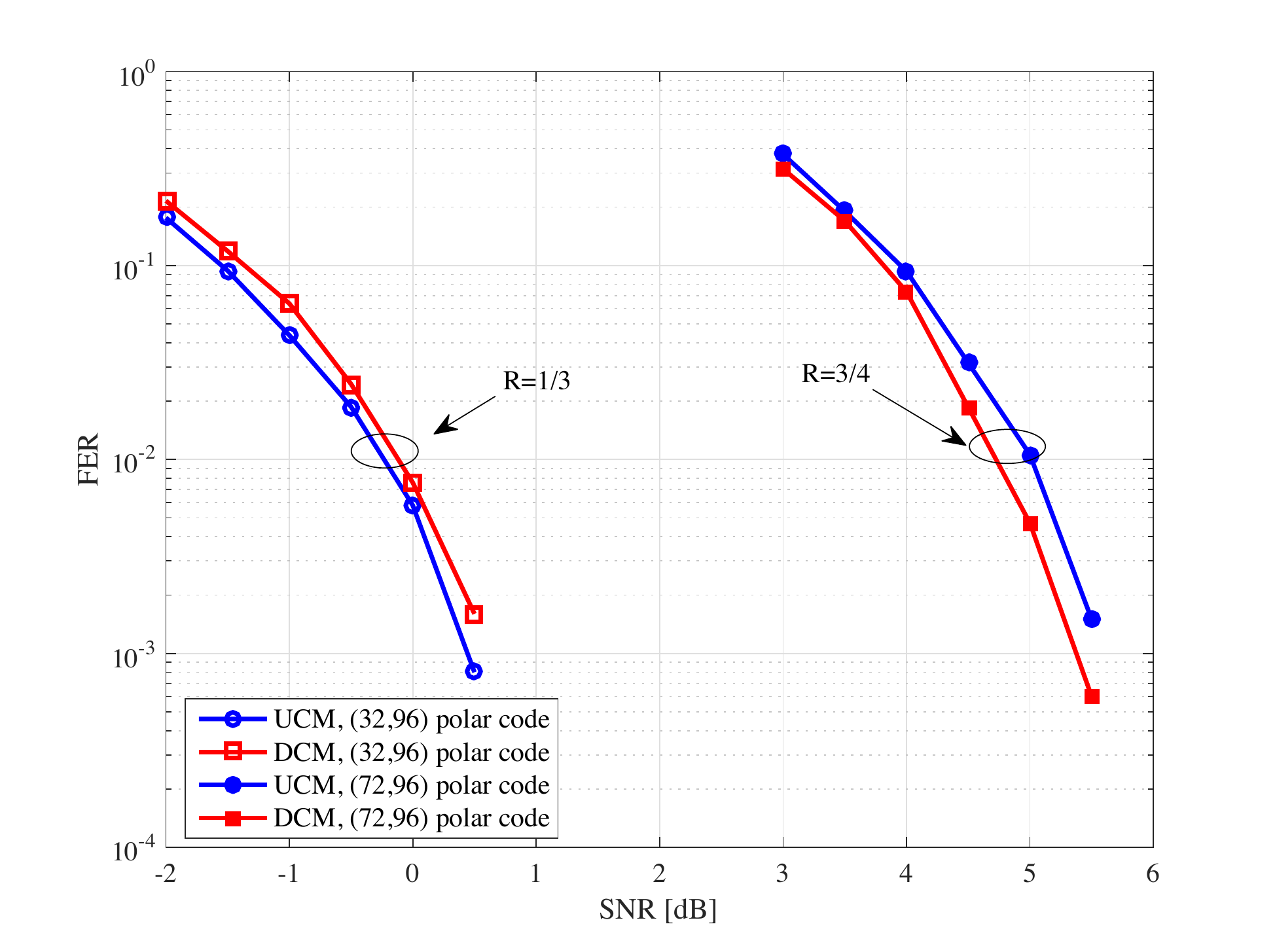}}
\caption{Performance comparisons of UCM and DCM approaches for various code rates. Here, the punctured polar codes of $N_p=96$ are obtained by puncturing $32$ bits from mother polar code of $N=128$.}
\label{fig1}
\end{figure}

%%%%%%%%%%%%%%%%%%%%%%%%%%%%%%%%%%%%%%%%%%%%%%%%%%%%%%%%%%
\section{Catastrophic Puncturing Patterns}\label{sec:CPP}

From the analysis in Section~\ref{sec:puncturing}, we learned that an information set $\Ac$ should be carefully designed for a punctured polar code, by taking into account a puncturing pattern and a chosen stochastic modelling. However, this approach cannot be used on constructing a rate-compatible code since an information set should not be changed according to the puncturing patterns in the family, i.e., a common information set, usually optimized for a mother code, should be kept. Motivated by this, we in this section analyze the impact of puncturing patterns for a fixed (pre-designed) information set $\Ac$.

%%%%%%%%%%%%%%%%%%%%%%%%%%%%%%%%%%%%%%%%%%%%%%%%%%%%%%%%%%%%%%%%%%%%%%%%%%%%%%%%%%
\begin{definition} A puncturing pattern $\pv^{N}$ is said to be catastrophic for a polarized channel $i$ (obtained from any underlying channel $W$) if
\begin{equation}\label{eq:boolean_fn}
Z^{(i)}(\pv^N)= 0.
\end{equation} More generally, a puncturing pattern $\pv^{N}$ is called  catastrophic for a set of polarized channels $i \in \Ac \subseteq [0:N-1]$ if
\begin{equation}
Z^{(i)}(\pv^N) = 0 \mbox{ for any } i \in \Ac.
\end{equation}
\end{definition}
\vspace{0.1cm}
Obviously, a catastrophic puncturing pattern is {\em universal} with respect to the underlying BI-DMC channel $W$. If a catastrophic puncturing pattern is used, then the block-error-rate (BLER) of the corresponding punctured code is always 1 and thus, it should be avoid. From the boolean functions in Section~\ref{sec:puncturing}, we obtain the two interesting facts:
\begin{itemize}
\item  Any puncturing pattern $\pv^{N}$ with $w_h(\pv^N) > N - 2^{wt(i)}$ is non-catastrophic for the polarized channel $i$.
\item  The boolean functions $Z^{(i)}(\pv^N)$ can be always be expressed in the corresponding {\em maximal} forms, in which each term represents a catastrophic puncturing pattern. 
\end{itemize}
In the example of $N=4$, the first fact is verified from  (\ref{eq:minimal1})-(\ref{eq:minimal4}). Also, the {\em maximal} form of $Z^{(2)}(\pv^4)$ is obtained using the fact $p_i \wedge \bar{p}_i = 0$ as
\begin{align*}
Z^{(2)}(\pv^4) =& p_{0}p_{2} \wedge p_{1}p_{3}\\
=&p_{0}p_{2}(p_{1}\wedge\bar{p}_{1})(\bar{p}_{3}\wedge p_{3})\wedge p_{1}p_{3}(p_{0}\wedge\bar{p}_{0})(\bar{p}_{2}\wedge p_{2})\\
=&p_{0}p_{1}p_{2}p_{3} \wedge p_{0}p_{1}p_{2}\bar{p}_{3} \wedge p_{0}\bar{p}_{1}p_{2}p_{3} \wedge p_{0}\bar{p}_{1}p_{2}\bar{p}_{3}\wedge p_{0}p_1\bar{p}_{2}p_{3} \wedge \bar{p}_{0}p_{1}p_{2}p_{3} \wedge \bar{p}_{0}p_1\bar{p}_{2}p_{3},
\end{align*} and the corresponding catastrophic puncturing patterns are 
\begin{align}\label{eq:example}
\{&(0,0,0,0), (0,0,0,1), (0,1,0,0),(0,1,0,1),(0,0,1,0),(1,0,0,0),(1,0,1,0)\}.
\end{align} Using the boolean function $Z^{(i)}(\pv^N)$, we can check if $\pv^N$ is a catastrophic or not for a given polarized channel $i$. Also, we can efficiently check it using the following lemma.

%%%%%%%%%%%% MODIFIED THEOREM %%%%%%%%%%%%%%%%%

\begin{lemma}\label{lem:cat-check} A puncturing pattern $\pv^{N}$ is  non-catastrophic for a polarized channel $i$ if the following {\em rank-increment} condition is satisfied:
\begin{align*}
&\mbox{Rank}\left(\Gm^N\left([i:N],\Bc_{\pv^{N}}^c\right)\right) - \mbox{Rank}\left(\Gm^{N}\left([i+1:N],\Bc_{\pv^{N}}^c\right) \right) = 1.
%&=^{(b)} w_{h}(\pv^{N}).
\end{align*}
\end{lemma}
\begin{IEEEproof}  When $\pv^{N}$ is applied, the capacity of the polarized channel $i$ is given by
\begin{align}
Z^{(i)}(\pv^{N}) &=  I(u_i; \{y_j: j \in \Bc_{\pv^N}^c\}|\uv_1^{i-1})\nonumber\\
&=H(u_i) - H(u_i| \uv^{N}\Gm\left([1:N],\Bc_{\pv^{N}}^c\right) ,\uv_1^{i-1})\nonumber\\
 &=H(u_i) - H(u_i| \{y_j: j \in \Bc_{\pv^N}^c\},\uv_1^{i-1})\nonumber\\
 &=1 - H\Big(u_i | u_{i}\Gm\left(i,\Bc_{\pv^{N}}^c\right) + \uv_{i+1}^{N}\Gm\left([i+1:N],\Bc_{\pv^{N}}^c\right)\Big),\nonumber
\end{align} where the interference term caused by the known values $\uv_1^{i-1}$, i.e., $ \uv_{1}^{i-1} \Gm([1:i-1],\Bc_{\pv^{N}}^c)$, is eliminated. Then, we have that
\begin{equation*}
 H\left(u_i| \{y_j: j \in \Bc_{\pv^N}^c\},\uv_1^{i-1}\right) = 0,
\end{equation*}  if $\Gm(i,\Bc_{\pv^{N}}^c)$ is linearly independent from the rows of $\Gm([i+1:N],\Bc_{\pv^{N}}^c)$.  This completes the proof.

\end{IEEEproof}

In Section~\ref{sec:Analysis}, we analyze the catastrophic puncturing patterns and then in Section~\ref{sec:SRP}, we present an efficient algorithm to construct a non-catastrophic puncturing pattern.

%%%%%%%%%%%%%%%%%%%%%%%%%%%%%%%%%%%%%%%%%%%%%%%%%%%%%%%%%%%%%%%%%%%%%%%%%%%%%%%%%%
\subsection{Analysis of Catastrophic Puncturing Patterns}\label{sec:Analysis}

For a given polarized channel $i$ with  $g(i)=(b_{n},\ldots,b_{1})$, we characterize all catastrophic puncturing patterns and their weight distributions.

%%%%%%%%%%%%%%%%%%%%%%%%%%%%%%%%%%%%%%%%%
\subsubsection{Characterization of catastrophic puncturing patterns}\label{subsec:CPP}

\begin{figure}
\centerline{\includegraphics[width=12cm]{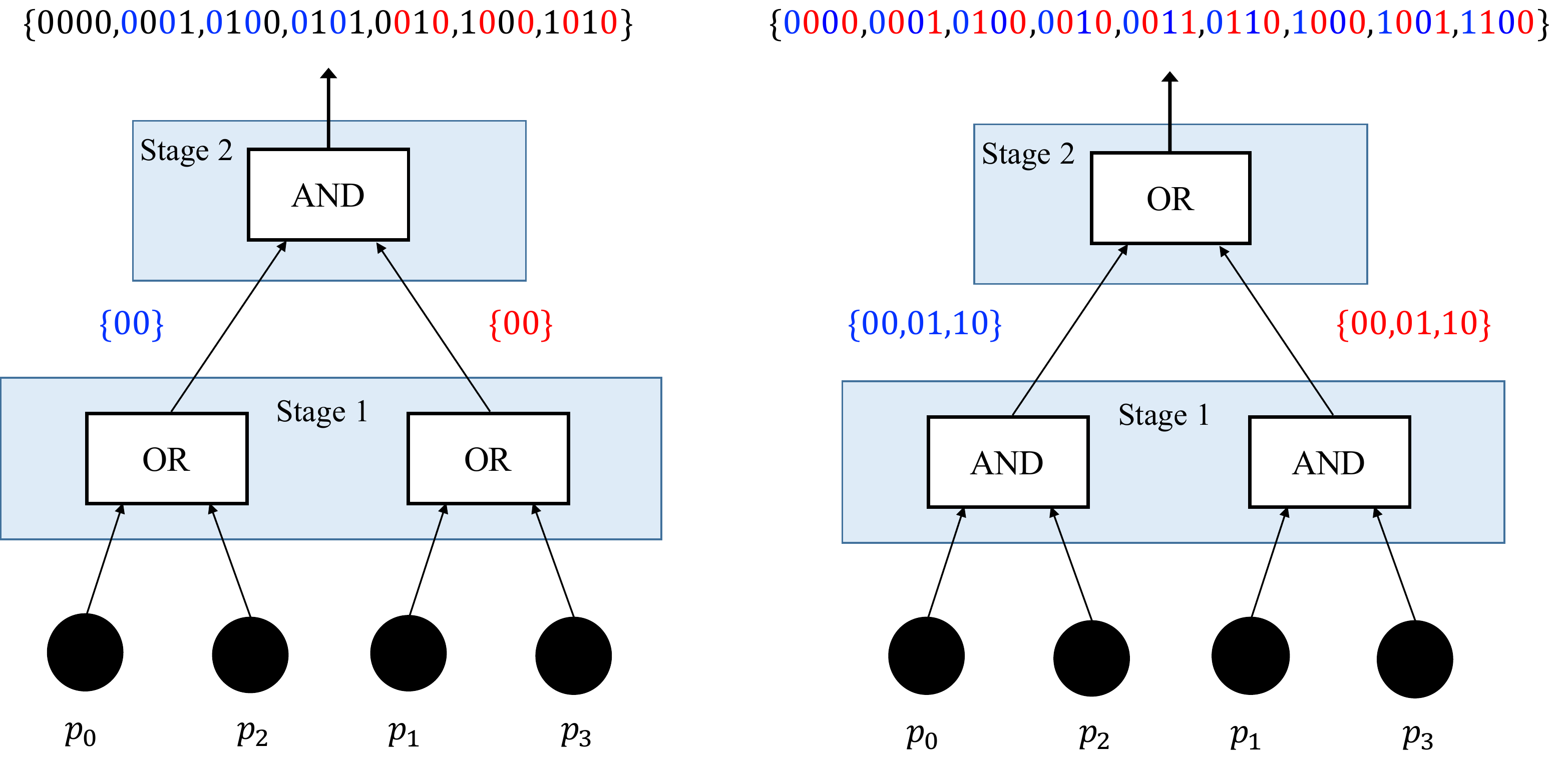}}
\caption{The binary trees to characterize the catastrophic puncturing patterns for $n=2$.}
\label{fig:Zset}
\end{figure}

Define $\Cc_{a}^{N}(b_{n},\ldots,b_{1})$ by the set of catastrophic puncturing patterns of the polarized channel $i$, i.e.,
\begin{equation}
\Cc_{a}^{N}(b_{n},\ldots,b_{1})\eqdef \{\pv^{N} \in \{0,1\}^N: Z^{(i)}(\pv^N)=0\}.
\end{equation} Now we develop an efficient recursive method to characterize $\Cc_{a}^{N}(b_{n},\ldots,b_{1})$. We first construct a {\em binary tree} consisting of $N=2^n$ leaf nodes and $n$ levels.   When $n=4$, there are four types of binary trees, two of which are illustrated in  Fig.~\ref{fig:Zset}. Note that the binary tree has two types of function nodes as ``AND" node and ``OR" node, and leaf nodes. Also, each  level $m$ contains the same types of function nodes in which the type is completely determined by the $b_m$  as
\begin{equation}
\left\{
  \begin{array}{ll}
    \mbox{AND node}, & b_m = 0 \\
    \mbox{OR node} & b_m = 1.
  \end{array}
\right.
\end{equation} 
%%%%%%%%%%%%%%%%%%%%%%%%%%%%%%%%%%%%%%%%%%%%%%%%%%%%%%%%%%%%%%%%%%%%%%%%%%%%%%%%%%%%
%%%%%%%%%%%%%%%%%%%%%%%%%%%%%%%%%%%%%%%%%%%%%%%%%%%%%%%%%%%%%%%%%%%%%%%%%%%%%%%%%%%%%%
For a given node $v$, we define the subtree that contains  all child nodes of $v$. As before, we can derive the boolean function associated with the subtree. Then, the output of the node  $v$ (i.e., the message value of its outgoing edge) represents the set of all binary sequences associated with the leaf nodes in the subtree such that each binary sequence in the set is the root of the boolean function. Due to the symmetry of the binary tree, all edges in the same level have the same message value (see Fig.~\ref{fig:Zset}).

We let $\Ic$ and $\Oc$ represent the incoming and outgoing message values at a node, respectively. From Fig.~\ref{fig:Zset}, the message value of the outgoing edge from $v$ is computed using the following update rules:

{\it $\bullet$ Update rule at OR node:}
\begin{equation}\label{eq:updateOR}
\Oc = \{(\pv_{1},\pv_{2}): \pv_{1}, \pv_{2} \in \Ic\}.
\end{equation}

{\it $\bullet$ Update rule at AND node:}
\begin{equation}\label{eq:updateAND}
\Oc=\{(\pv_{1} \bigtriangledown \pv_{2}): \pv_{1}, \pv_{2} \in \Ic\},
\end{equation} where, for the ease of expression, we used the notation of
\begin{align*}
&\{(\pv_{1} \bigtriangledown \pv_{2}): \pv_{1}, \pv_{2} \in \Ic\} \eqdef \{(\pv_{1},\pv_{2}): \pv_{1} \in \Ic, \pv_{2} \in \{0,1\}^{l_\Ic}  \}\\
&\;\;\;\;\;\;\;\;\;\;\;\;\;\;\;\;\;\;\;\;\;\;\;\;\;\;\;\; \cup \{(\pv_{1},\pv_{2}): \pv_{1} \in \{0,1\}^{l_\Ic}, \pv_{2} \in \Ic\},
\end{align*}where $l_\Ic$ denote the length of binary sequences in $\Ic$. 
Using the binary tree and the above update rules, we can find all catastrophic puncturing patterns recursively. Starting with $n=1$, we have:
\begin{equation}
\Cc_{a}^{2}(0) = \{00,01,10\} \mbox{ and } \Cc_{a}^{2}(1) = \{00\}.
\end{equation} Using them, we obtain $\Cc_{a}^{4}(b_{2},b_{1})$  as
\begin{equation*}
\Cc_{a}^{4}(b_{2},b_{1})= \left\{
                \begin{array}{ll}
                  \{(\pv_{1},\pv_{2}): \pv_{1} \bigtriangledown \pv_{2} \in \Cc_{a}^{2}(b_{2})\}, & b_{1}=0 \\
                  \{(\pv_{1},\pv_{2}): \pv_{1},\pv_{2} \in \Cc_{a}^{2}(b_{2})\}, &b_{1}=1.
                \end{array}
              \right.
\end{equation*} Repeatedly applying the above procedures, we have:

\begin{proposition} For a polarized channel $i$ with $g(i) =(b_{n},\ldots,b_1)$, the set of catastrophic puncturing patterns are obtained recursively as
\begin{align*}
&\Cc_{a}^{2^{n-m}}(b_{n},\ldots,b_{m+1}) =\\
& \left\{
                \begin{array}{ll}
                  \{(\pv_{1} \bigtriangledown \pv_{2}): \pv_{1} , \pv_{2} \in \Cc_{a}^{2^{n-m-1}}(b_{n},\ldots,b_{m+2}), &  b_{m+1}=0 \\
                  \{(\pv_{1},\pv_{2}): \pv_{1},\pv_{2} \in \Cc_{a}^{2^{n-m-1}}(b_{n},\ldots,b_{m+2}), & b_{m+1}=1
                \end{array}
              \right.
\end{align*} for $m=n-2,\ldots,1,0$, where the recursion starts with
\begin{equation*}
\Cc_{a}^{2}(0) = \{00,01,10\} \mbox{ and } \Cc_{a}^{2}(1) = \{00\}.
\end{equation*}
\end{proposition}
\vspace{0.1cm}

%%%%%%%%%%%%%%%%%%%%%%%%%%%%%%%%%%
\subsubsection{Weight distributions}

We derive the weight distributions of catastrophic puncturing patterns recursively, in which the recursion starts with simple polynomial of $z$ at each coded bit location. First of all, for any two input polynomial $D_{\alpha}(z) = \sum_{s=0}^{m} d_{\alpha}(s)z^{s}$ and $D_{\beta}(z) = \sum_{s=0}^{n} d_{\beta}(s)z^{s}$, where $d_{\alpha}(s)$ and $d_{\beta}(s)$ denote the corresponding numbers of catastrophic puncturing patterns with zero weight $s$ (i.e. $s$ zeros), we obtain the output polynomials after OR and AND operations as:

{\it $\bullet$ OR operation:}
\begin{equation}
f_{1}(D_{\alpha}(z),D_{\beta}(z)) \eqdef D_{\alpha}(z)D_{\beta}(z).\label{eq:OR}
\end{equation}

{\it $\bullet$ AND operation:}
\begin{align}
 f_{0}(\alpha(z),\beta(z)) &\eqdef D_{\alpha}(z)(1+z)^m + D_{\beta}(z)(1+z)^n  - D_{\alpha}(z)D_{\beta}(z),\label{eq:AND}
\end{align}where $m$ and $n$ denote the highest degrees of $D_{\alpha}(z)$ and $D_{\beta}(z)$, respectively. Note that (\ref{eq:OR}) and (\ref{eq:AND}) are derived from (\ref{eq:updateOR}) and (\ref{eq:updateAND}), respectively. As performed in Section~\ref{subsec:CPP}, using the binary tree and the above update rules, we  get:

%%%%%%%%
\begin{proposition}\label{thm:dist} The weight distribution of the set of catastrophic puncturing patterns for a polarized channel $i$ of a length-$N$ polar code can be computed recursively by
\begin{align}
D_{i}^{(N)}(z) &\eqdef \sum_{s=2^{w_h(i)}}^{N} d_i^{(N)}(s)z^s \nonumber\\
& = f_{[i]\!\!\!\!\!\mod\!2}\left(D_{\lfloor i/2\rfloor}^{(N/2)}(z),D_{\lfloor i/2\rfloor}^{(N/2)}(z)\right),\label{eq:weight}
\end{align}for $i \in \{0,1,2...,N-1\}$, where $D_{0}^{(1)}(z) \eqdef z$.
\end{proposition}
\vspace{0.1cm}
For example, from Theorem~\ref{thm:dist}, we can derive the 
\begin{align}
D_{2}^{(4)}(z) &= 2z^2 + 4z^3 + z^4,\label{eq:dist_ex}
\end{align} which is well-matched to  (\ref{eq:example}). 

%From the minimum degrees of the weight distributions, we can see which polarized channels are more robust to a puncturing, which can be used to design an information set that is robust to a puncturing.  From Lemma~\ref{lem:degree} below, we can easily compute the minimum degree of a weight distribution.

%%%%%%%%%%%%%%%%
%\begin{lemma}\label{lem:degree} The minimum degree of its weight distribution $D_{i}^{(N)}(z)$ is equal to $2^{w_h(g(i))}=2^{wt(i)}$.
%\end{lemma}
%\begin{IEEEproof} See the longer version of this paper.
%\end{IEEEproof}

%%%%%%%%%%%%%%%%%%%%%%%%%%%%%%%%%%%%%%%%%%%%%%%%
\section{Construction of Rate-Compatible Polar Code}\label{sec:RC}

In this section, we will construct a RC polar code to send $k$ information bits with $M$ various code rates $R_1 > R_2 > \cdots > R_M$, where a puncturing pattern $\pv_{i}^{N}$ is applied to a mother polar code of length $N=k/R_M$ to achieve a target rate $R_i$. For all code rates, a fixed information set $\Ac$ is used. Also, for the simplicity, it is assumed that $N$ has the form of power of 2.  Then, we present the two efficient methods, named greedy and reciprocal constructions, to construct RC polar codes where each puncturing pattern 
$\pv_{i}^N$ in the family is non-catastrophic.

\begin{algorithm}[h]
\caption{Greedy algorithm to find $\pv_{0}^{N}$ }\label{GA}
\begin{algorithmic}[1]
\State Initialization:
\State $\Ac_{\rm o}=\{i_{1},\ldots,i_{k}\}$ denotes an ordered information set of $\Ac$ with respect to Hamming weight.
\State $\pv_{0}^{N}= (0,\ldots,0)$, $\Bc_{\pv_{0}^{N}}=\phi$, and $t=1$.
%\Procedure{GA}{$\Ac'$}\Comment{Input: an ordered information set $\Ac'$ of size $k$}
\While{$t\leq k$}
\For{$j=1,\ldots,N: j  \notin \Bc_{\pv_{0}^{N}} $}
\If{ $Z^{(i_{t})}(\pv_{0}^{N}+\ev_j) = 1$}
\State $\pv_{0}^{N} =\pv_{0}^{N} + \ev_j$ 
\newline\Comment{$\ev_j$ denotes an unit vector with 1 in the $j$th location.}
\State $t=t+1$ and \textbf{go to} 4.
\ElsIf{j=N}
\State $\pv_{0}^{N} = \pv_{0}^{N} + \ev_r$
\newline\Comment{$r$ is randomly and uniformly chosen from the zero locations in $\pv_{0}^{N}$}
\State $t=t+1$
\EndIf
\EndFor
\EndWhile
%\EndProcedure
\end{algorithmic}
%Note that the condition in line 6 can be efficiently checked by using the rank-increment in Lemma~\ref{lem:cat-check}.
\end{algorithm}

%%%%%%%%%%%%%%%%%
\subsection{Greedy Construction}\label{subsec:GC}

The main idea of the greedy construction is as follows:

\begin{enumerate}
\item Find a possibly {\em minimum} weight non-catastrophic $\pv_{0}^{N}$ such that
\begin{equation}\label{eq:alg}
Z^{(i)}(\pv_{0}^{N}) = 1 \mbox{ for all } i \in \Ac.
\end{equation} Note that if $w_{\rm H}(\pv_{0}^{N})>w_{\rm H}(\pv_{1}^{N})$, then it is impossible to construct the desired RC polar code. In this section, it is assumed that $w_{\rm H}(\pv_{0}^{N}) \leq w_{\rm H}(\pv_{1}^{N})$.
\item To obtain a non-catastrophic puncturing pattern $\pv_{1}^{N}$, we add $w_{\rm H}(\pv_{1}^{N}) - w_{\rm H}(\pv_{0}^{N})$ ones to  some zero locations in $\pv_{0}^{N}$.
\item Next, to obtain a non-catastrophic puncturing pattern $\pv_{2}^{N}$, we add $w_{\rm H}(\pv_{2}^{N}) - w_{\rm H}(\pv_{1}^{N})$ ones to  some zero locations in $\pv_{1}^{N}$.
\item In general, to obtain a non-catastrophic puncturing pattern $\pv_{i}^{N}$, we add $w_{\rm H}(\pv_{i}^{N}) - w_{\rm H}(\pv_{i-1}^{N})$ ones to  some zero locations in $\pv_{\rm 1}^{N}$.
\item Repeat the above procedures until obtaining all the puncturing patterns in the family.
\end{enumerate} Note that in the above, the locations to add ones are chosen arbitrary (i.e., randomly and uniformly, or via a clever algorithm). Then, the proposed algorithm is established by developing a greedy algorithm to find the $\pv_{0}^{N}$ (see Algorithm~\ref{GA}).
In this algorithm, we used an {\em ordered} information set $\Ac_{\rm o}=\{i_1,\ldots,i_k\}$ (with respect to a Hamming weight) from $\Ac$, where $w_{\rm H}(g(i_{j})) \geq w_{\rm H}(g(i_{k}))$ for any $j, k \in \Ac_{\rm o}$ with $j \geq k$. This is motivated by the fact that a polarized channel $i$ tends to have more non-catastrophic puncturing patterns as $w_{\rm H}(g(i))$ becomes larger. Thus, the use of the $\Ac_{\rm o}$ makes it easier to satisfy the  condition in line 6 in Algorithm 1, by adding fewer ones' in $\pv_{0}^{N}$. Eventually, we are able to obtain a lower weight $\pv_{0}^{N}$ to satisfy (\ref{eq:alg}). Note that Algorithm 1 is a randomized algorithm since, when the condition in line 6 in Algorithm is not satisfied, i.e., we need to add more ones to $\pv_{0}^{N}$ to satisfy the condition, one is added to an arbitrary zero location of $\pv_{0}^{N}$. Note that although the use of $\Ac_{\rm o}$ is helpful to find a lower weight 
 $\pv_{0}^{N}$, it does not ensure that Algorithm 1 finds the minimum weight $\pv_{0}^N$.

\subsection{Reciprocal Construction}\label{subsec:RC}

In this section, we construct a family of rate-compatible puncturing patterns $\pv_{1}^{N},...,\pv_{M-1}^N$ where they are reciprocal. We first focus on the design of $\pv_{M-1}^{N}=(p_{M-1,0},...,p_{M-1,N-1})$. In the proposed approach, the $|\Bc_{\pv_{M-1}}^N|$ number of zero locations will be chosen so that the one-covering property in Theorem~\ref{thm1} is satisfied. Then, we design the $\pv_{M-2}^N$ by adding the $|\pv_{M-2}^{N}|-|\pv_{M-1}^N|$ number of additional zero locations to the $\pv_{M-1}^{N}$. This process will be continued until constructing all the puncturing patterns in the family. The specific construction methods will be explained as follows. 
Define the useful set operation below:
\begin{definition} For any $i, j \in [0:N-1]$, we say that $i \perp j$ if $ w_{\rm H}(g(i)\oplus g(j)) = w_{\rm H}(g(i))+w_{\rm H}(g(j))$, where $\oplus$ denotes a binary addition.   Then, for any two index sets $\Mc$ and $\Nc$, define a set operation as
\begin{equation*}
\Mc \boxplus \Nc = \left\{g^{-1}\left(g(i) \oplus g(j)\right): i \perp j, i\in \Nc, j\in \Mc \right\}.
\end{equation*} 
%Also, for any two ordered sets $\Ac=\{i_1,...,i_m\}$ and $\Bc=\{j_1,...,j_n\}$, we let 
 %$\cup_{\rm o}$ denote the {\em ordered union} as
%\begin{align*}
%\Ac \cup_{\rm o} \Bc &=\{i_1,...,i_m,j_1,...,j_n\}\\
%\Bc \cup_{\rm o} \Ac &=\{j_1,...,j_n,i_1,...,i_m\}.
%\end{align*} 
\end{definition}
\vspace{0.2cm}
Using the above definitions and a given information set $\Ac$, we derive the ordered sets $\Lc_{\Ac}^{(i)}, i=0,...,n$, in a recursive manner as
\begin{align*}
\Lc_{\Ac}^{(0)} &= \{0\}\eqdef\{i^{(0)}_1\}\\
\Lc_{\Ac}^{(1)} &= \{i \in [0:N-1]: w_{\rm H}(g(i)) = 1 \} \setminus \Ac \eqdef \{i^{(1)}_1,...,i^{(1)}_{|\Lc_{\Ac}^{(1)}|}\}\\
\Lc_{\Ac}^{(j)} &=  \left(\Lc_{\Ac}^{(j-1)}\boxplus \Lc_{\Ac}^{(1)}\right)\setminus \Ac \eqdef \{i^{(j)}_1,...,i^{(j)}_{|\Lc_{\Ac}^{(j)}|}\},
\end{align*} for $j=2,...,n$. Then, we define a sequence of length $\sum_{j=1}^{n}|\Lc_{\Ac}^{(j)}$ as
\begin{equation}
\Ic_{\rm seed}(\Ac) = \left(i^{(0)}_1,i^{(1)}_1,...,i^{(1)}_{|\Lc_{\Ac}^{(1)}|},...,i^{(n)}_1,...,i^{(n)}_{|\Lc_{\Ac}^{(n)}|}\right),
\end{equation} where this sequence depends on the choice of an information set.
Leveraging this sequence, we construct the rate-compatible puncturing patterns (equivalently, $\Bc_{\pv_{i}^N}$) as follows:
\begin{enumerate}
\item The $\Bc_{\pv_{M-1}^N}$ is constructed by taking the first $|\Bc_{\pv_{M-1}^N}|$ elements of $\Ic_{\rm seed}(\Ac)$. By construction of 
 $\Lc_{\Ac}^{(i)}$'s, the corresponding puncturing pattern  $\pv_{M-1}^N$ is reciprocal and non-catastrophic with respect to $\Ac$. 
\item Next, the $\Bc_{\pv_{M-2}^N}$ is constructed by taking the first $|\Bc_{\pv_{M-2}^N}|$ elements of $\Ic_{\rm seed}(\Ac)$. Clearly, $\Bc_{\pv_{M-1}^N}\subset \Bc_{\pv_{M-2}^N}$, namely, rate-compatibility is satisfied.
\item In general,  the $\Bc_{\pv_{i}^{N}}$ is constructed by taking the first $|\Bc_{\pv_{i}^N}|$ elements of $\Ic_{\rm seed}(\Ac)$.
\end{enumerate} Note that if $|\Ic_{\rm seed}(\Ac)|<|\Bc_{\pv_{1}^N}|$, then it is impossible to construct the desired RC polar code. In this case, we need to change the highest code rate. In the following example, we show that the supportable highest rate is determined as a function of an information set.

%%%%%%%%%%%%%%%%%%%%%%%%%%%%%%%%%%%%%
\begin{example} Suppose that $N=8$ (e.g., $n=3$) and $\Ac=\{4,6\}$. Then we have:
 \begin{align*}
 \Lc_{\Ac}^{(0)} &= \{0\} \mbox{ and } \Lc_{\Ac}^{(1)} = \{1,2\}\\
 \Lc_{\Ac}^{(2)} &= (\Lc_{\Ac}^{(1)} \boxplus \Lc_{\Ac}^{(1)}) \setminus \Ac =  \{g^{-1}((0,0,1)\oplus (0,1,0))\}=\{3\}\\
 \Lc_{\Ac}^{(3)} &=(\Lc_{\Ac}^{(2)} \boxplus \Lc_{\Ac}^{(1)}) \setminus \Ac = \phi.
 \end{align*} Then, we have $\Ic_{\rm seed}(\{4,6\})=(0,1,2,3)$. Also, in other example of $\Ac=\{5,7\}$, we have:
 \begin{align*}
 \Lc_{\Ac}^{(0)} &= \{0\} \mbox{ and } \Lc_{\Ac}^{(1)} = \{1,2,4\}\\
 \Lc_{\Ac}^{(2)} &= (\Lc_{\Ac}^{(1)} \boxplus \Lc_{\Ac}^{(1)}) \setminus \Ac =\{3,6\}\\
 \Lc_{\Ac}^{(3)} &=(\Lc_{\Ac}^{(2)} \boxplus \Lc_{\Ac}^{(1)}) \setminus \Ac = \phi.
 \end{align*} Then, we have $\Ic_{\rm seed}(\{5,7\}) = (0,1,2,4,3,6)$. This example shows that  the number of possible punctured bits varies according to the choice of information set $\Ac$. 
 \end{example}

%%%%%%%%%%%%%%%%%%%%%%%%%%%%%%%%%%%%%%%%%%%%%%%%%%%%%%%%%%%%%%%%%%%%%%%%%%%%%%%
%%%%%%%%%%%%%%%%%%%%%
\begin{figure}
\centerline{\includegraphics[width=14cm]{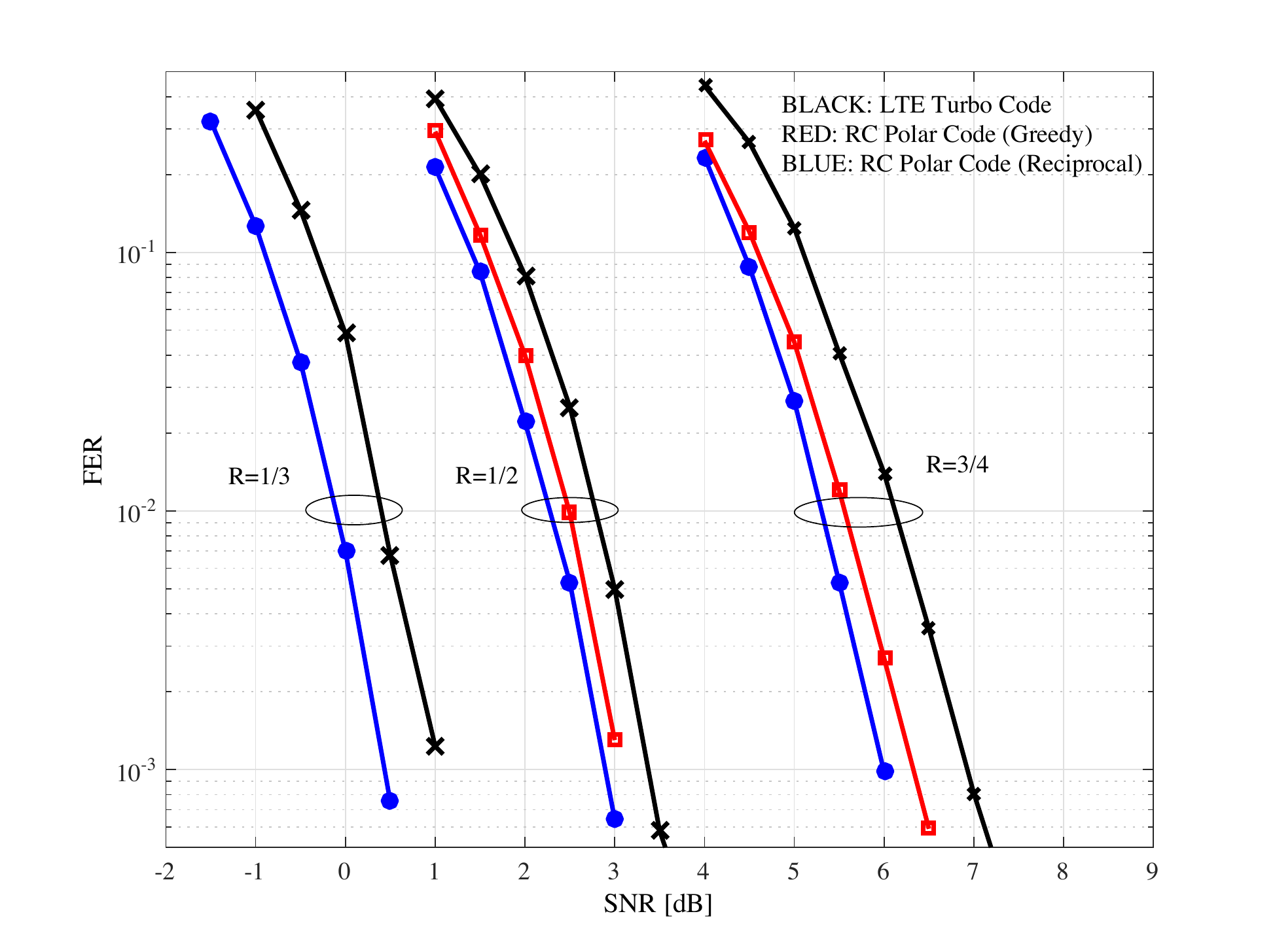}}
\caption{Performance comparisons of RC polar codes and RC Turbo codes. Black lines denotes the performances of the RC Turbo code. Also, red and blue lines denote the performances of the RC polar codes constructed by greedy and reciprocal constructions, respectively.}
\label{fig2}
\end{figure}

%%%%%%%%%%%%%%%%%%%%%
\section{Numerical Results}\label{sec:NR}

In this section, we evaluate the performances of the proposed puncturing methods. For the simulation, AWGN channel and BPSK modulation are assumed.

{\bf Length-flexibility:} In this case, a puncturing is performed to construct a polar code whose code length is not the power of 2. 
We employed a polar code with $N=128$. To construct the punctured polar code of $N_p = 96$, we use the QUP and RQUP in Section~\ref{subsec:RPP} for UCM and DCM, respectively. Also, CRC-aided list decoder is used with 8-bit CRC and list-size 8. Fig.~\ref{fig1} shows that UCM  performs better than DCM at rate 1/3, while vice versa, at rate 3/4. Hence, we can learn that either UCM or DCM should be carefully chosen according to the code rate of a target punctured polar code.

{\bf Rate-compatible polar code:} Fig.~\ref{fig2} shows the performance comparisons of the proposed RC polar codes and RC Turbo code. For the simulation, we constructed the two RC polar codes where where one is developed by greedy construction in Section~\ref{subsec:GC} and the other is by reciprocal construction in Section~\ref{subsec:RC}.  As mother codes, both polar and Turbo codes have the length of 256 (e.g., $N=256$). Also, for polar code, CRC-aided list decoder is used with 5-bit CRC and list-size 32. Hence, the number of information bits for polar code is 93 while that of Turbo code is 88. From this simulation, we observe that the proposed RC polar codes outperform the RC Turbo code adopted in LTE. Also, we can see that a low-complexity reciprocal construction performs very well.

%We further construct a rate-compatible puncturing pattern in which, using Algorithm 1, we first find a non-catastrophic puncturing pattern to obtain a rate $3/4$ and then, add extra ones to this puncturing pattern to obtain a rate 1/2 where the locations are randomly and uniformly chosen. We compare the performance of the proposed rate-compatible punctured polar code with rate-compatible Turbo code used in LTE. Fig.~\ref{fig2} shows that the proposed rate-compatible punctured polar code outperforms the rate-compatible LTE Turbo code.

%%%%%%%%%%%%%%%%%%%%%%%%%%%%%%%%%%%%%%%%%%%%%%%%%%%%%%%%%%%%%%
%\begin{algorithm}[h]
%\caption{One-covering puncturing patterns}\label{GA}
%\begin{algorithmic}[1]
%\State Initialization: 
%\State $\Lc_{\Ac}^{(0)} = \{0\}$ and  $\Lc_{\Ac}^{(1)} = \{\mbox{all weight-1 locations}\} \setminus \Ac$
%\Procedure{GA}{$\Ac'$}\Comment{Input: an ordered information set $\Ac'$ of size $k$}
%\For{$j=2,\ldots,n$}
%\State $\Lc_{\Ac}^{(j)} = \left(\Lc_{\Ac}^{(j-1)}\boxplus \Lc_{\Ac}^{(j-1)}\right)\setminus \Ac$.
%\EndFor
%\end{algorithmic}
%Note that the condition in line 6 can be efficiently checked by using the rank-increment in Lemma~\ref{lem:cat-check}.
%\end{algorithm}

%%%%%%%%%%%%%%%%%%%%
\section{Conclusion}\label{sec:con}

We derived the boolean expressions of the capacities of polarized channels of finite-length polar codes. Based on this, we provided a guideline to jointly optimize a puncturing and the corresponding information set, Also, it was shown that for each fixed information set, there exist the catastrophic puncturing patterns that should be avoided to yield a good performance. Furthermore, we presented two efficient methods to construct non-catastrophic puncturing patterns for any fixed information set. Leveraging them, we designed the RC polar codes using non-catastrophic puncturing patterns. Via simulation results, it was demonstrated that the proposed RC polar codes can outperform the RC Turbo code adopted in LTE. Therefore, polar codes can be a good candidate for 5G channel coding.

%\section*{Acknowledgement}

%This work was supported in part by the Electronics and Telecommunications Research Institute through the Korean Government (Wireless Transmission Technology in Multi-point to Multi-point Communications) under Grant 16ZI1100.

%%%%%%%%%%%%%%%%%%%%%%%%%%%%%%%%%%%%%%%%%%%

\end{document}